\documentclass[journal]{IEEEtran}
\usepackage{tabularx}
\usepackage{cite}
\input epsf
\usepackage{graphicx}
\usepackage{hyperref}
\usepackage{graphicx}
\usepackage{amssymb}
\usepackage{amsmath}
\usepackage[super]{nth}
\usepackage{fixltx2e}

\begin{document}
\title{\LARGE All-Back-Schottky-Contact Thin-Film Photovoltaics}
	
\author{
	\IEEEauthorblockN{Marco Nardone\IEEEauthorrefmark{1}} \\
	\IEEEauthorblockA{\IEEEauthorrefmark{1}Bowling Green State University, Bowling Green, OH 43403, USA}}

\maketitle

\begin{abstract}
\boldmath
The concept of All-Back-Schottky-Contact (ABSC) thin-film photovoltaic devices is introduced and evaluated using 2D numerical simulation.  The built-in field is generated by reach-through Schottky junctions with two metals of different work functions in an alternating, side-by-side pattern along the back of the device.  This approach minimizes the number of interfaces through which charge transport is required while eliminating the need for transparent conductive front contacts, pn junctions, and scribing.  Performance metrics are calculated for a range of design parameters and it is shown that passivation of the front surface and back contacts is critical to achieve efficiencies near 20\% using only one layer of material deposited on the back contacts.  This design paradigm may afford a wide range of passivation materials since both electrical conduction \emph{and} optical transparency would not be required of any one passivation layer.
\end{abstract}

\section{Introduction}\label{sec:intro}

Thin film photovoltaic (TFPV) technologies have the distinct advantages of high-throughput deposition methods and material flexibility. Although TFPV laboratory devices based on various materials have recently achieved greater than 20\% efficiency\cite{green2015}, further advancements are required to tighten the gap between cell and module performance, reduce cost, and improve long-term reliability.  Conventional TFPV cells are comprised of several stacked layers of semiconductors and dielectrics sandwiched between electrodes, with a transparent conducting oxide (TCO) acting as the electrode on the illuminated side.  Other layers are required to form pn-junctions and passivation regions which improve performance.  However, with each layer comes interfaces which always carry with them high densities of recombination centers, energy barriers, and seeds of degradation\cite{wolden2011} . Performance limiting effects are also spawned by the unique physics of inherently nonuniform films that are on the order of 1 $\mu$m in thickness\cite{karpov2002}.

\begin{figure*}[htb]\centering
\includegraphics[width=0.80\textwidth]{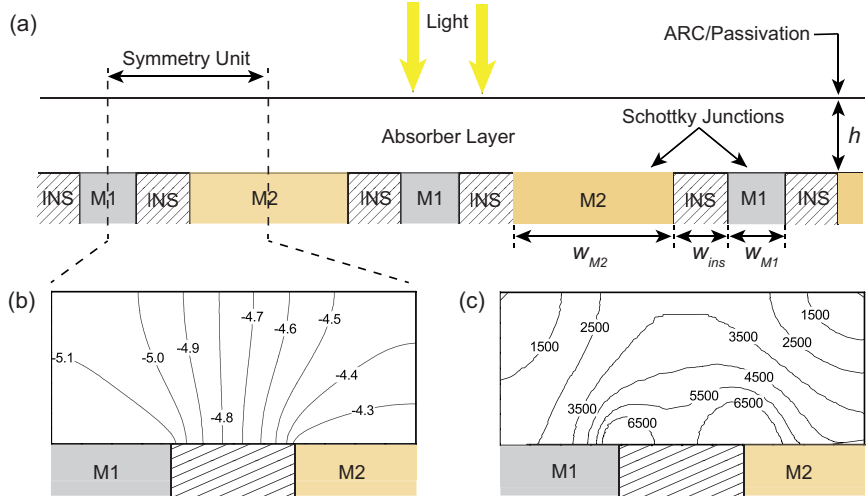}
\caption{ABSC thin-film device: (a) schematic with semiconductor absorber of thickness $h$ atop alternating layers of metal 1 (M1), insulator (INS), and metal 2 (M2); (b) electric potential distribution in volts relative to vacuum with metal work functions of $\phi_{M1}=4.2$ V, $\phi_{M2}=5.2$ V (other parameters are provided in Table \ref{tbl:param}) with $w_{M1}/2=w_{M2}/2=w_{ins}=h=1$ $\mu$m.  (c) electric field magnitude distribution in V/cm for the parameters of (b).\label{Fig:schematic}}
\end{figure*}

As an alternative to the conventional stacked approach, this work uses numerical simulation to explore the concept of an All-Back-Schottky-Contact (ABSC) TFPV device architecture wherein two dissimilar metals of different work functions are adhered to the non-illuminated side of the device.  Reach-through Schottky junctions and the difference in work functions of the two metals establish the built-in field.  The basic philosophy of this design paradigm is to minimize the number of interfaces through which charge transport is required.  A schematic layout is shown in Fig. \ref{Fig:schematic}(a).

Interdigitated back contact (IBC) crystalline silicon (c-Si) solar cells were first conceived\cite{lammert1977} in the 1970's and high efficiency modules are now commercially available\cite{masuko2014, smith2012}.  Several groups continue to study this technology which typically employs n-type c-Si wafer material of 100--300 $\mu$m in thickness with alternating p-type and n-type zones along the back to affect charge separation and collection\cite{mingirulli2011, lee2014, tomasi2014, lu2007, castano2011, woehl2015}. Thin-film Si devices (2--10 $\mu$m thick) have also been developed with efficiencies of greater than 10\% \cite{haschke2014,keevers2007,jeong2013}. All of these back contacted approaches employ pn-junctions of some type.

Recently, lateral organic heterojunction cells were developed with IBCs of alternating metals\cite{kim2015}.  It was shown that such a contacting scheme can in fact produce a built-in field that effectively separates charge carriers if the metals are of sufficiently different work functions.  In that work, the back contact materials were Al, Au, and MoO$_3$-coated Au with measured work functions of 4.1, 5.1, and 5.6 eV respectively.  That work also demonstrated the feasibility of using UV-lithography to create metal lines down to 5 $\mu$m wide separated by a channel as narrow as 200 nm.  Here we investigate how such a contacting scheme may be used to fabricate high-efficiency TFPV cells with absorber layers comprised of materials similar to polycrystalline CdTe or CIGS.

The important distinction of ABSC cells from conventional TFPV is that the reach-through Schottky junctions can generate an electric field of $E>10^3$ V/cm throughout the absorber [see Fig. \ref{Fig:schematic}(c)], thereby enabling effective electron-hole separation without the need for complex doping steps. Some of the main advantages of the ABSC design include: (1) lack of TCO, pn-junctions, and window layers, thus reducing parasitic photon absorption and interface issues; (2) no scribing requirements for module fabrication thereby reducing damage and shadow losses; and (3) opportunities for design optimization via optical control and back contact shape/size.  Several of those points could also help to reduce losses in going from cell to module.  On the other hand, simplification of the semiconductor components leads to added complexity in the metal contacting scheme, which is among several possible challenges, such as, fabricating a high density of Schottky contacts, shunting between contacts, and material compatibility between metals of different work functions and the semiconductor.

\section{Physical Characteristics}\label{sec:physical}

\begin{figure}[htb]\centering
	\includegraphics[width=0.45\textwidth]{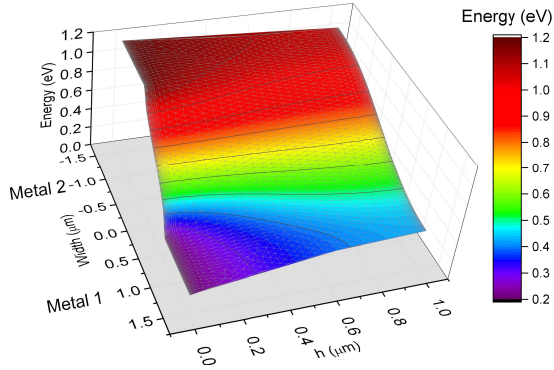}
	\caption{2D conduction band in eV relative to the Fermi level ($E=0$ plane) at equilibrium.  The back of the device is at $h=0$.\label{Fig:3D}}
\end{figure}

The basic physics of ABSC-TFPV is similar to that of metal-semiconductor-metal (MSM) structures, but with unique features due to the device architecture.  For example, the electric potential varies in 2D, such that the standard 1D energy band diagram is insufficient and the 2D bands follow the contours in Fig. \ref{Fig:schematic}(b), where it can be seen that both transverse and lateral electric fields are generated, with the magnitudes indicated in Fig. \ref{Fig:schematic}(c).  The calculated conduction band (CB) surface is illustrated in Fig. \ref{Fig:3D} where the same parameter values were used as in Fig. \ref{Fig:schematic}. The steep CB gradient between the metals (lateral direction) and the relatively small gradient between the back and front surfaces (transvers direction) are evident in Fig. \ref{Fig:3D}.
  
Stacked devices are always modeled with the directions of light propagation and carrier currents parallel to each other.  That is not the case in ABSC devices, however, and a 2D analysis is required to understand the unique interplay between drift and diffusion currents relative to the direction of light propagation.  There are several characterstic lengths to consider (in what follows, we assume a p-type semiconductor and refer to the variables shown in Fig. \ref{Fig:schematic}).  First, the depletion width, $W_D$ of the Schottky contact (which we assume to be ideal) at M2 should be within the absorber thickness; $h<W_D$.  The depletion width is given by, \cite{sze2007}
\begin{equation}\label{eqn:depletion}
W_D=\sqrt{\frac{2\epsilon_0\epsilon_r\psi_{bi}}{q N_A}},
\end{equation}
where $q$ is the elementary charge, $\psi_{bi}$ is the built-in potential, $\epsilon_0$ is the permittivity of free space, $\epsilon_r$ is the relative permittivity of the semiconductor, and $N_A$ is the acceptor concentration.  In addition, the semiconductor must be thick enough to absorb the majority of incident photons; $h>\alpha^{-1}$ where $\alpha$ is the absorption coefficient of the semiconductor.  Thus the conditions,
\begin{equation}\label{eqn:h}
\alpha^{-1}<h<W_D,
\end{equation}
provide the optimal range for the thickness.  The depletion width also plays a role in determining the lateral field between the metals which are separated by the dielectric material of width $w_{ins}$.  If $\Delta\phi_M$ is the difference in metal work functions, a field on the order of $E\approx \Delta\phi_M/w_{ins}$ will be generated between the metals when,
\begin{equation}
w_{ins}<W_D,
\end{equation}
such that the field is not screened by space charges.

Another important condition is related to the diffusion length of minority carriers, $L_n=\sqrt{D\tau}$, because they must migrate from the quasi-neutral zone above M1 to the high-field/depleted region near the insulator and, finally, to M2. Therefore,
\begin{equation}\label{eqn:m1}
w_{M1}\lesssim 2\sqrt{\mu\tau kT/q},
\end{equation}
where the Einstein relation was used for the diffusion coefficient, $D=\mu kT/q$ with thermal voltage $kT/q$, minority carrier mobility $\mu$, and $\tau$ the minority carrier lifetime.  The factor of 2 in Eq. (\ref{eqn:m1}) accounts for the requirement of carrier migration for half of the contact width.  The conditions on the width of metal M2 appear to be less stringent since it will depend mostly on the majority carrier diffusion length $L_p$; hence,
\begin{equation}\label{eqn:m2}
w_{M2}\lesssim 2L_p.
\end{equation}
The electric field distribution above M2 will also affect minority carrier collection and, therefore, determine how $w_{M2}$ will affect performance.  Front surface conditions (charge accumulation, front surface fields, etc.) will strongly affect the field above M2 (discussed further below).

\section{Numerical Simulation Results}\label{sec:numerical}

In what follows, power conversion efficiency ($\eta$), open circuit voltage ($V_{oc}$), short-circuit current ($J_{sc}$), and fill factor ($FF$) were determined using numerical simulation for a range of parameters including contact widths ($w_{M1}$ and $w_{M2}$), absorber layer thickness ($h$), surface recombination velocity at the back contacts ($S_{bc}$) and at the front surface ($S_{fs}$).  Fixed values of $w_{ins}=1$ $\mu$m and the difference between metal work functions of $\Delta\phi_{M}=1$ eV were used along with the parameter set given in Table \ref{tbl:param}; rather generic for p-type TFPV such as CdTe,\cite{gloeckler2003} CIGS,\cite{gloeckler2003}, CZTS,\cite{patel2012}, and perovskites.\cite{liu2014}  Shockley-Read-Hall (SRH) recombination was assumed in the bulk of the absorber with donor-type defects located at mid-gap.  Also, ideal Schottky contacts were assumed for this initial study.

Performance metrics were calculated using COMSOL Multiphysics with a 2D geometry.  Simulations were performed both in the dark and with exposure to AM1.5G spectrum at 100 mW/cm$^2$ intensity.  The photo-generation rates for this study simply followed the Beer Law of exponential photon absorption with a direct optical absorption co-efficient\cite{fahrenbruch1983} [$\alpha(\lambda)=A\sqrt{h\nu-E_g}$, with $A=10^5$ cm$^{-1}$ eV$^{-1/2}$, photon energy $h\nu$, and band gap $E_g$], neglecting reflections at the front and back layers.  More rigorous optical calculations, such as light trapping techniques could be included using the same simulation platform.

\begin{table}[htb]\centering
	\centerline { TABLE I  } 
	\vskip5pt
	\centerline {\textsc{Generic Parameter Values.}}
	\vskip2pt
	\begin{tabular}{c c c c} \hline\hline
		Parameter & Unit & Value & Description\\
		\hline
		$E_g$        & eV & 1.5 & band gap\\
		$\chi$       & eV & 4.0 & electron affinity\\
		$\epsilon_r$ & [0] & 10 & relative permittivity\\
		$N_c$        & cm$^{-3}$ & $10^{18}$ & effective CB DOS \\
		$N_v$        & cm$^{-3}$ & $10^{19}$ & effective VB DOS\\
		$\mu_n$      & cm$^2$ V$^{-1}$ s$^{-1}$ & 100 & electron mobility\\
		$\mu_p$      & cm$^2$ V$^{-1}$ s$^{-1}$ & 10  & hole mobility\\
		$N_a$        & cm$^{-3}$  & $10^{14}$ & acceptor doping\\
		$N_t$		 & cm$^{-3}$  & $10^{14}$ & defect concentration\\
		$E_t$		 & eV         & 0.75      & defect energy above VB\\
		$\sigma_n$	 & cm$^2$     & 10$^{-12}$ & electron cross-section\\
		$\sigma_p$	 & cm$^2$     & 10$^{-15}$ & hole cross-section\\
		$\phi_{M1}$	 & eV  & 4.2 & metal M1 work function\\
		$\phi_{M2}$	 & eV  & 5.2 & metal M2 work function\\
		\hline
	\end{tabular}
	\\[1.5mm]
	Note: CB, VB, and DOS stand for conduction band, valence band, and density of states, respectively.
	\end{table}
\refstepcounter{table}\label{tbl:param}

\begin{figure}[htb]\centering
\includegraphics[width=0.45\textwidth]{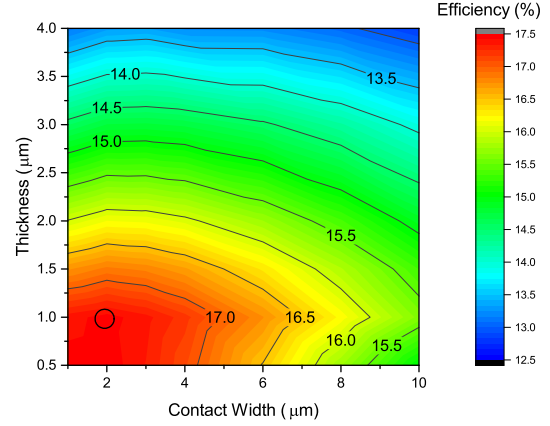}
\caption{Efficiency contours as a function of absorber layer thickness and contact widths (assuming $w_{M1}=w_{M2}$) and the parameters of Table \ref{tbl:param} with front surface and back contact recombination velocities of $S_{fs}=0$ and $S_{bc}=100$ cm/s, respectively.  Parameters corresponding to the circled area at $w_{M1}=w_{M2}=2$ are further evaluated in Figs. \ref{Fig:metrics} and \ref{Fig:metricsSfs}.\label{Fig:eff1}}
\end{figure}

We first consider optimistic surface recombination by setting $S_{fs}=0$ and $S_{bc}=100$ cm/s.  Bulk SRH recombination is significant with a minority carrier lifetime of $\tau=(\sigma_n N_t v_{th})^{-1}=1$ ns, using the parameters in Table \ref{tbl:param} and thermal velocity $v_{th}=10^7$ cm/s.  As shown in Fig. \ref{Fig:eff1}, the efficiency calculations within the parameter space of $0.5<h<4$ $\mu$m and $1<w_{M1}=w_{M2}<10$ $\mu$m were within the range of $13\% <\eta<18$\% with the largest value given by an optimum thickness of $h\approx 1$ $\mu$m, which satisfies Eq. (\ref{eqn:h}); for our parameter values, $W_D\approx 3$ $\mu$m and $\alpha^{-1}\approx 0.1$ $\mu$m. Also, narrower contact width $w_{M1}$ results in better performance according to Eq. (\ref{eqn:m1}), given our minority carrier diffusion length of $L_n\approx 0.5$ $\mu$m.  

\begin{figure}[htb]\centering
	\includegraphics[width=0.45\textwidth]{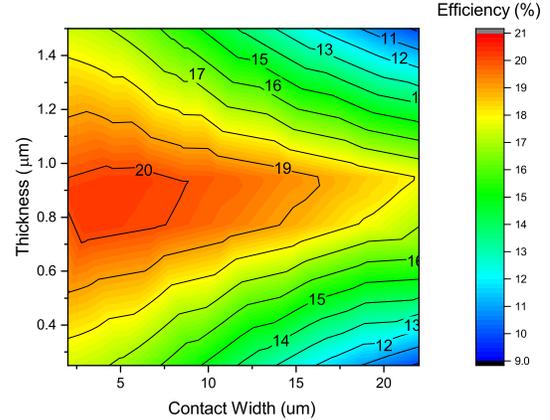}
	\caption{Efficiency contours as a function of absorber layer thickness and contact widths (assuming $w_{M1}=w_{M2}$) with the same parameters as Fig. \ref{Fig:eff1} except for an increase of minority carrier lifetime from $\tau=1$ ns to $\tau=10$ ns.\label{Fig:eff2}}
\end{figure}

Fig. \ref{Fig:eff2} shows the resulting increase in efficiency when the minority carrier lifetime is increased to $\tau=10$ ns (by lowering the mid-gap defect concentration to $N_t=10^{13}$ cm$^{-3}$).  The efficiency spans from 9\% to 21\% over a range of contact widths $2<w_{M1}=w_{M2}<22$ $\mu$m and absorber thicknesses $0.25<h<1.5$ $\mu$m.  In this case, efficiency remains above 18\% for up to 20-$\mu$m contact width when the absorber thickness is optimized at $h\lesssim 1$.  According to Eq. (\ref{eqn:m1}) the allowable contact width increases relative to the case with $\tau=1$ ns shown in Fig. \ref{Fig:eff1}.  The optimum thickness decreased slightly below 1 $\mu$m because the lower density of occupied defect states affects space charge and slightly lowers the built-in field which, in turn, lowers $W_D$ in Eq. (\ref{eqn:depletion}). 

It should be noted that Eq. \ref{eqn:depletion} may be an oversimplification of the depletion width because $W_D$ will also depend on: (i) the compensation of $N_A$ by charged defect states; and (ii) the insulating boundary condition at the front surface which requires that the normal component of the field satisfies $\mathbf{n}\cdot \epsilon \mathbf{E}=Q_t$, where $\mathbf{n}$ is the surface normal unit vector and $Q_t$ is the surface charge density due to traps there.  As stated below, we have assumed neutral surface traps ($Q_t=0$) and, therefore, a flat electric potential at the front boundary which reduces the field throughout the entire absorber thickness above M2.  The properties of both the bulk and surface defects play important roles in determining the field distribution $\mathbf{E}(\mathbf{x})$.

\begin{figure}[htb]\centering
	\includegraphics[width=0.35\textwidth]{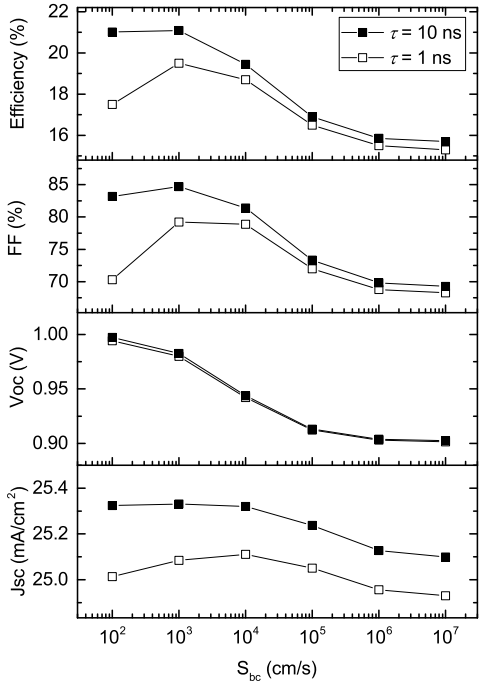}
	\caption{Performance metrics vs. back contact surface recombination velocity $S_{bc}$ assuming $S_{fs}=0$ at the front surface and minority carrier lifetimes of $\tau=1$ ns (parameters from Table \ref{tbl:param}) and $\tau=10$ ns. $w_{M1}=w_{M2}=2$ $\mu$m and $h=1$ $\mu$m.\label{Fig:metrics}}
\end{figure}

Fig. \ref{Fig:metrics} shows the effects of $S_{bc}$ on the performance metrics assuming the dimensions of the circled point in Fig. \ref{Fig:eff1} and minority carrier lifetimes of $\tau=1$ ns and $\tau=10$ ns. Maximum $\eta$ of 19.5\% and 21.5\% were achieved at $S_{bc}=10^3$ cm/s, which then decayed to nearly stable values between 15.5\% and 16\% when $S_{bc}=10^7$ cm/s (this scenario uses $S_{fs}=0$ cm/s).  It is apparent that $S_{bc}$ affected efficiency mainly through $FF$ and $V_{oc}$.  The latter monotonically decayed from 1.0 V to 0.9 V while $FF$ displayed large, non-monotonic variations that correlated with $\eta$.  The non-montonic behavior of $\eta$ and $FF$ was due to the effect of $S_{bc}$ on the minority carrier distribution and, therefore, on the electric field (as is the case in a high-injection scenario, $n\geq N_a$).  Our calculations show that increasing $S_{bc}$ decreases the transverse field at the metal contacts while increasing the lateral field above the insulator, resulting in an optimum $S_{bc}$.

\begin{figure}[htb]\centering
	\includegraphics[width=0.35\textwidth]{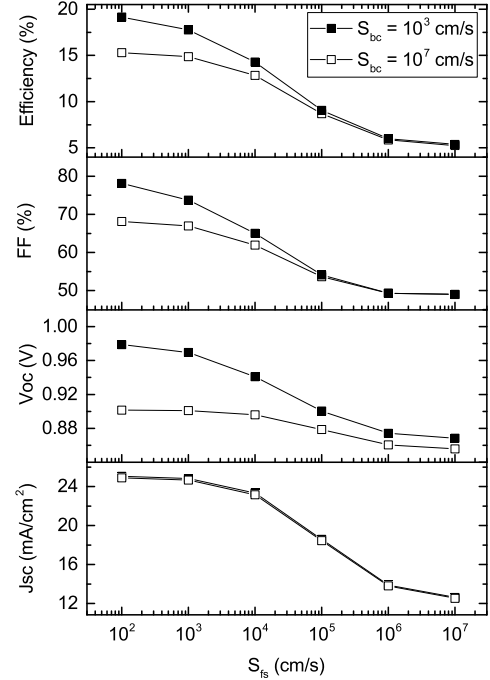}
	\caption{Performance metrics vs. front surface recombination velocity $S_{fs}$ for $S_{bc}=10^7$ cm/s and $S_{bc}=10^3$ cm/s at the back contacts.  Device parameters are from Table \ref{tbl:param} with $\tau=1$ ns, $w_{M1}=w_{M2}=2$ $\mu$m, and $h=1$ $\mu$m.\label{Fig:metricsSfs}}
\end{figure}

Next we consider front surface effects as governed by $S_{fs}=\sigma_s v_{th} N_{st}$.  We set equal capture cross-sections for electrons and holes, $\sigma_s=10^{-14}$ cm$^2$, $v_{th}=10^7$ cm/s, and allowed the surface trap density to vary in the range $10^9\geq N_{st}\geq 10^{14}$ cm$^{-2}$ so that $10^2 \geq S_{fs}\geq 10^7$ cm/s.  As an initial test, we assumed only neutral centers located at mid-gap.  We also considered both the worse case scenario of $S_{bc}=10^7$ cm/s and passivated $S_{bc}=10^3$ cm/s.  The profound effect of front surface recombination is evident in Fig. \ref{Fig:metricsSfs} as $\eta$ drops from 15.3\% to 5.3\% over the range of $S_{fs}$ for $S_{bc}=10^7$ cm/s.  Improving the back surface passivation so that $S_{bc}=10^3$ cm/s increased $\eta$ to a maximum of 19.1\% but resulted in no performance gain at large values of $S_{fs}$, which dominated efficiency reduction.  Losses were greatest for $FF$ and $J_{sc}$, while $V_{oc}$ exhibited a relatively small loss.  Qualitatively, those variations are a result of the increase in recombination in the area of greatest photo-absorption (at the front) while the Schottky junction field is strongest near the back.

\section{Discussion}\label{sec:discussion}

From Figs. \ref{Fig:metrics} and \ref{Fig:metricsSfs} it is clear that the challenges of dealing with interface quality still remain, but the ABSC cell allows for passivation of the front surface by materials that need not carry photo-current (i.e. high resistivity and optically transparent).  Conventional wisdom suggests that a PV cell should have the greatest photo-absorption in the region of strongest electric field to optimize electron-hole pair separation.  Our approach goes against that logic but we have shown that a high performance device may still be possible due to the presence of both lateral and transverse built-in electric fields. An added benefit of the ABSC design is that the Schottky junctions are buried and therefore protected against the degrading effects of high carrier injection.\cite{nardone2014}

Experimental efforts to develop a proof of concept is of foremost importance but additional theoretical analysis may include a broader exploration of the parameter space, optimal band gap grading, light trapping measures, low level light concentration, and a front surface field layer.  Such measures could improve efficiency while making the absorber layer thinner and the contacts wider.  Also, the simulation work presented here did not include the resistance of the metal contacts, which must be considered when specific designs are studied.

\section{Conclusions}\label{sec:conclusions}

In summary, an alternative design paradigm of all-back-Schottky-contact TFPV has been introduced.  The numerical analysis results suggest that this approach may lead to simple yet high efficiency, durable TFPV devices if surface recombination can be effectively passivated. In addition, we have demonstrated the use of COMSOL Multiphysics$^\circledR$ as a numerical simulation platform for 2D modeling of TFPV devices.

\bibliographystyle{IEEEtran}
\bibliography{Tcell}


\end{document}